\documentclass[preprint,amsmath,showpacs]{revtex4}
\usepackage{graphicx}
\usepackage{dcolumn}
\usepackage{bm}
\usepackage{epstopdf}

\def\bee{\begin{eqnarray}}
\def\eee{\end{eqnarray}}

\begin{document}

\title{ Flavor Transition Mechanisms of Propagating Astrophysical Neutrinos -A Model
Independent Parametrization }
\author{Kwang-Chang Lai, Guey-Lin Lin and T. C. Liu}\affiliation{Institute of Physics, National
Chiao-Tung University, Hsinchu 300, Taiwan}\affiliation{Leung Center
for Cosmology and Particle Astrophysics, National Taiwan University,
Taipei 106, Taiwan.}

\date{\today}

\begin{abstract}

One of the important goals for future neutrino telescopes is to
identify the flavors of astrophysical neutrinos and therefore
determine the flavor ratio. The flavor ratio of astrophysical
neutrinos observed on the Earth depends on both the initial flavor
ratio at the source and flavor transitions taking place during
propagations of these neutrinos. We propose a model independent
parametrization for describing the above flavor transitions. A few
flavor transition models are employed to test our parametrization.
The observational test for flavor transition mechanisms through our
parametrization is discussed.

\end{abstract}

 \maketitle Recent developments
of neutrino telescopes
\cite{Berghaus:2008bk,km3net,Gorham:2008yk,Collaboration:2009uy,Allison:2009rz}
have inspired numerous efforts of studying neutrino flavor
transitions utilizing astrophysical neutrinos as the beam source
\cite{Beacom:2002vi,Barenboim:2003jm,Beacom:2003nh,Beacom:2003zg,Pakvasa:2004hu,Costantini:2004ap,Bhattacharjee:2005nh,
Serpico:2005sz,Serpico:2005bs,Xing:2006uk,Winter:2006ce,Xing:2006xd,
Majumdar:2006px,Rodejohann:2006qq,Meloni:2006gv,Blum:2007ie,Hwang:2007na,
Pakvasa:2007dc,Choubey:2008di,Maltoni:2008jr}. Given the same
neutrino flavor ratio at the source, some flavor transition models
predict rather different neutrino flavor ratios on the Earth
compared to those predicted by the standard neutrino oscillations
\cite{Pakvasa:2004hu}. In this article, we propose a scheme to
parametrize flavor transition mechanisms of astrophysical neutrinos
propagating from the source to the Earth. As will be shown later,
such a parametrization is very convenient for classifying flavor
transition models which can be tested by future neutrino telescopes.

To test flavor transition mechanisms, it is necessary to measure the
flavor ratio of astrophysical neutrinos reaching to the Earth. The
possibility for such a measurement in IceCube has been discussed in
Ref.~\cite{Beacom:2003nh}. It is demonstrated that the $\nu_e$
fraction can be extracted from the measurement of the muon track to
shower ratio by assuming flavor independence of the neutrino
spectrum and the equality of $\nu_{\mu}$ and $\nu_{\tau}$ fluxes on
the Earth due to the approximate $\nu_{\mu}-\nu_{\tau}$ symmetry
\cite{Balantekin:1999dx,Harrison:2002et}. Taking a neutrino source
with fluxes of $\nu_{e}$ and $\nu_{\mu}$ given by $E_{\nu_e}^2{\mbox
d}N_{\nu_e}/{\mbox d}E_{\nu_e}=0.5E_{\nu_{\mu}}^2{\mbox
d}N_{\nu_{\mu}}/{\mbox d}E_{\nu_{\mu}}=10^{-7}$ GeV cm$^{-2}$
s$^{-1}$, which is roughly the order of the Waxman-Bahcall bound
\cite{Waxman:1998yy}, and thresholds for muon and shower energies
taken to be $100$ GeV and $1$ TeV, respectively, the $\nu_e$
fraction can be determined to an accuracy of $25\%$ at IceCube for
$1$ yr of data taking, or equivalently to an accuracy of $8\%$ for a
decade of data taking. Such an accuracy is obtained for a $\nu_e$
fraction in the vicinity of $1/3$. The accuracies corresponding to
other central values of the $\nu_e$ fraction are also presented in
Ref.~\cite{Beacom:2003nh}. The $\nu_{\mu}$ to $\nu_{\tau}$ event
ratio can also be measured in IceCube. However, the accuracy of this
measurement is limited by the low statistics of $\nu_{\tau}$ events.

\begin{figure}[htb]
\includegraphics[scale=0.33]{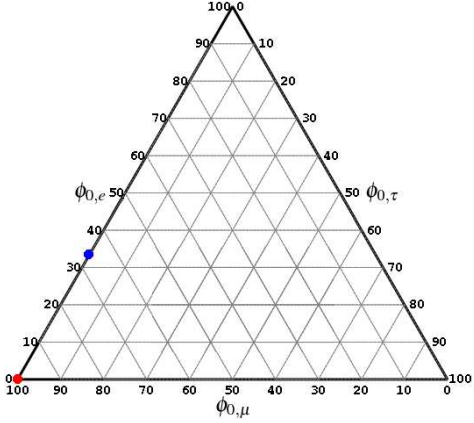}
\caption{The ternary plot for representing the flavor ratio of
astrophysical neutrinos. The numbers on each side of the triangle
denote the flux fraction of a specific flavor of neutrino. The blue
point, situated on the left side of the triangle, marks the pion
source $\Phi_{0,\pi}=(1/3,2/3,0)^T$ and the red point, situated at
the lower-left corner of the triangle, marks the muon-damped source
$\Phi_{0,\mu}=(0,1,0)^T$. } \label{example}
\end{figure}

 {\it Neutrino flavor ratio represented by the ternary plot.}--
To study neutrino flavor transitions, we describe the neutrino
flavor composition at the source by a normalized flux
$\Phi_0=(\phi_{0,e},~\phi_{0,\mu},~\phi_{0,\tau})^T$ satisfying the
condition \cite{remark}
\begin{eqnarray}
&&\phi_{0,e}+\phi_{0,\mu}+\phi_{0,\tau}=1, \nonumber \\
&&\phi_{0,\alpha}\geq0, \hspace{.5mm}{\rm for}\hspace{1mm}
\alpha=e,\mu,\tau,\label{norma}
\end{eqnarray}
where each $\phi_{0,\alpha}$ is the sum of neutrino and antineutrino
fluxes. Any point on or inside the triangle shown in
Fig.~\ref{example} represents a specific flavor ratio characterizing
the source.
The triangular region bounded by vertices $(1,~0,~0)^T$,
$(0,~1,~0)^T$, and $(0,~0,~1)^T$ contains all possible source types
in terms of flavor ratios. The pion source and the muon-damped
source with flavor compositions $\Phi_{0,\pi}=(1/3,2/3,0)^T$ and
$\Phi_{0,\mu}=(0,1,0)^T$, respectively, are explicitly marked on the
figure \cite{sources}.

The net effect of flavor transition processes occurring between the
source and the Earth is represented by the matrix $P$ such that
\begin{equation}
\Phi=P\Phi_0, \label{P_representation}
\end{equation}
where $\Phi=(\phi_{e},~\phi_{\mu},~\phi_{\tau})^T$ is the flux of
neutrinos reaching to the Earth. We note that our convention implies
$P_{\alpha\beta}\equiv P(\nu_{\beta}\to \nu_{\alpha})$.

{\it $Q$ matrix parametrization for flavor transitions of
astrophysical neutrinos.}-- Since the triangular region in
Fig.~\ref{example} represents all possible neutrino flavor
composition at the source, it is convenient to parametrize $\Phi_0$
by \cite{Lai:2009ke}
\begin{equation}
\Phi_0=\frac{1}{3}{\mathbf V}_1+a{\mathbf V}_2+b{\mathbf V}_3,
\label{source_flux}
\end{equation}
where ${\mathbf V}_1=(1,1,1)^T$, ${\mathbf V}_2=(0,-1,1)^T$, and
${\mathbf V}_3=(2,-1,-1)^T$. Mathematically, ${\mathbf V}_1/3$
represents the center of the triangle, while $a{\mathbf V}_2$ and
$b{\mathbf V}_3$ represent horizontal and vertical displacements
within the triangle, respectively. The ranges for $a$ and $b$ are
$-1/3+b\leq a\leq 1/3-b$ and $-1/6\leq b\leq 1/3$ such that
Eq.~(\ref{source_flux}) covers all points of the triangular region.
The pion source and the muon-damped source mentioned in
Fig.~\ref{example} correspond to $(a,b)=(-1/3,0)$ and
$(a,b)=(-1/2,-1/6)$ respectively. In general, a source with a
negligible $\nu_{\tau}$ flux corresponds to $a=-1/3+b$.

The above parametrization for $\Phi_0$ is also physically motivated.
The vector ${\mathbf V}_1$ gives the normalization for the neutrino
flux since the sum of components in ${\mathbf V}_1/3$ is equal to
unity, while the sum of components in ${\mathbf V}_2$ and that in
${\mathbf V}_3$ are both equal to zero. The vector $a{\bf V}_2$
determines the difference between $\nu_{\mu}$ and $\nu_{\tau}$ flux,
$\phi_{0,\mu}-\phi_{0,\tau}$, while preserving their sum,
$\phi_{0,\mu}+\phi_{0,\tau}$. Finally the vector $b{\mathbf V}_3$
determines the sum of  $\nu_{\mu}$ and $\nu_{\tau}$ flux,
$\phi_{0,\mu}+\phi_{0,\tau}$, while preserving their difference
$\phi_{0,\mu}-\phi_{0,\tau}$.

Following the same parametrization, we write the neutrino flux
reaching to the Earth as
\begin{equation}
\Phi=\kappa {\bf V}_1+\rho {\bf V}_2+\lambda {\bf V}_3.
\label{measured_flux}
\end{equation}
It is easy to show that
\begin{equation}
\left(
  \begin{array}{c}
     \kappa \\
    \rho \\
    \lambda \\
  \end{array}
\right) = \left(
  \begin{array}{ccc}
    Q_{11} & Q_{12} & Q_{13} \\
     Q_{21} & Q_{22} & Q_{23} \\
     Q_{31} & Q_{32} & Q_{33} \\
  \end{array}
\right)
\left(
  \begin{array}{c}
     1/3 \\
    a \\
    b \\
  \end{array}
\right)
 , \label{new_basis}
\end{equation}
where $Q={\mathbf A}^{-1}P{\mathbf A}$ with
\begin{equation}
{\mathbf A}= \left(
  \begin{array}{ccc}
    1 & 0 & 2 \\
     1 & -1 & -1 \\
     1 & 1 & -1 \\
  \end{array}
\right). \label{evector}
\end{equation}
In other words, $Q$ is related to $P$ by a similarity transformation
where columns of the transformation matrix ${\mathbf A}$ correspond
to vectors ${\mathbf V}_1$, ${\mathbf V}_2$, and ${\mathbf V}_3$,
respectively.

The parameters $\kappa$, $\rho$ and $\lambda$ are related to the
flux of each neutrino flavor by
\begin{equation}
\phi_e=\kappa+2\lambda, \ \phi_{\mu}=\kappa-\rho-\lambda, \
\phi_{\tau}=\kappa+\rho-\lambda, \label{phi_krl}
\end{equation}
with the normalization $\phi_e+\phi_{\mu}+\phi_{\tau}=3\kappa$.
Since we have chosen the normalization
$\phi_{0,e}+\phi_{0,\mu}+\phi_{0,\tau}=1$ for the neutrino flux at
the source, the conservation of total neutrino flux during
propagations corresponds to $\kappa=1/3$. In general flavor
transition models, $\kappa$ could be less than $1/3$ as a
consequence of (ordinary) neutrino decaying into invisible states or
oscillating into sterile neutrinos. To continue our discussions, it
is helpful to rewrite Eq.~(\ref{phi_krl}) as
\begin{equation}
\rho=\left(\phi_{\tau}-\phi_{\mu}\right)/2, \
\lambda=\phi_e/3-\left(\phi_{\mu}+\phi_{\tau}\right)/6.
\label{measure_rl}
\end{equation}
It is then clear from Eqs.~(\ref{new_basis}) and (\ref{measure_rl})
that, for fixed $a$ and $b$, the first row of matrix $Q$ determines
the normalization for the total neutrino flux reaching to the Earth,
the second row of $Q$ determines the breaking of
$\nu_{\mu}-\nu_{\tau}$ symmetry in the arrival neutrino flux, and
the third row of $Q$ determines the flux difference
$\phi_e-(\phi_{\mu}+\phi_{\tau})/2$.

Compared to $P$, matrix $Q$ is very convenient for classifying
flavor transition models. First of all, those models which preserve
the total neutrino flux are characterized by the condition
$\sum_{\alpha=e,\mu,\tau} P_{\alpha\beta}=1$ in the $P$ matrix
parametrization. On the other hand, these flux-conserving models
must give $\kappa=1/3$ in the $Q$ matrix parametrization,
irrespective of the initial flavor composition characterized by
parameters $a$ and $b$. This implies $Q_{11}=1$ and
$Q_{12}=Q_{13}=0$ from Eq.~(\ref{new_basis}). Clearly the
flux-conservation condition in the $Q$ matrix parametrization is
much simpler. Second, for those models which do not seriously break
the $\nu_{\mu}-\nu_{\tau}$ symmetry, the second and third rows of
$P$ are almost identical, i.e., $(P_{\mu
e},P_{\mu\mu},P_{\mu\tau})\approx (P_{\tau
e},P_{\tau\mu},P_{\tau\tau})$, and the second and third columns of
$P$ are also almost identical, i.e., $(P_{e
\mu},P_{\mu\mu},P_{\tau\mu})^T \approx (P_{e
\tau},P_{\mu\tau},P_{\tau\tau})^T$. Using these conditions and the
relation $Q={\mathbf A}^{-1}P{\mathbf A}$, one can show that
$(Q_{21},Q_{22},Q_{23})\approx (0,0,0)$ and
$(Q_{12},Q_{22},Q_{32})^T\approx (0,0,0)^T$. Obviously, the
approximate $\nu_{\mu}-\nu_{\tau}$ symmetry is realized by a much
simpler condition in the $Q$ matrix parametrization. In summary, we
have seen that the first and second rows of $Q$ as well as the
matrix element $Q_{32}$ are already constrained in a simple way by
assuming the conservation of total neutrino flux and the validity of
approximate $\nu_{\mu}-\nu_{\tau}$ symmetry. Hence, under these two
assumptions, one can simply use the values for $Q_{31}$ and $Q_{33}$
to classify flavor transition models. This is the most important
advantage of $Q$ matrix parametrization. In fact, as will be
elaborated later, this parametrization is also very useful for
discussing the effect of flux nonconservation, i.e., the case with
$\kappa \neq 1/3$.

It was first discussed in Ref.~\cite{Barenboim:2003jm} that the
flavor measurement in neutrino telescopes is useful for studying
neutrino flavor composition at the astrophysical source (for recent
studies, see Refs.~\cite{Lai:2009ke,initial_ratio}) and neutrino
flavor transition properties during its propagation from the source
to the Earth (see also Ref.~\cite{Beacom:2002vi}). For probing
flavor transition properties of astrophysical neutrinos, the authors
of Ref.~\cite{Barenboim:2003jm} considered typical astrophysical
sources and applied the flavor transition matrix $P$ (denoted by
$\chi$ in the original paper) derived from the standard neutrino
oscillation model or flavor transition models involving new physics
for obtaining possible flavor ratios to be measured by terrestrial
neutrino telescopes. It was pointed out that there are some flavor
transition models which can produce rather distinctive neutrino
flavor ratios on the Earth compared to those produced by the
standard neutrino oscillation model, even with uncertainties of
neutrino mixing parameters taken into account. Hence these flavor
transition models can be tested on the basis of their flavor-ratio
predictions for astrophysical neutrinos arriving on the Earth. In
our approach, we test the fundamental structure of a given flavor
transition model, namely the $Q$ matrix of the model. As noted
earlier, possible neutrino flavor transition models which conserve
the total neutrino flux are encompassed by possible values of
$Q_{2i}$ and $Q_{3i}$ with $i=1,\, 2$, and $3$. In the
$\nu_{\mu}-\nu_{\tau}$ symmetry limit, only the values of $Q_{31}$
and $Q_{33}$ are relevant. The matrix elements of $Q$ can be
determined by performing fittings to the flavor-ratio measurements
in the neutrino telescopes, as will be demonstrated later. The
obtained ranges for these matrix elements can be used as the basis
for testing any flavor transition model.

 {\it Examples.}-- In the previous section, we have discussed the
 properties of the $Q$ matrix and their advantages.
 In this section, we shall confirm such properties using a few flavor
 transition models as examples. We begin by considering the
standard three-flavor neutrino oscillations. It is well known that
\begin{equation}
P^{\rm osc}_{\alpha\beta}=\sum_{i=1}^{3} |U_{\beta i}|^2|U_{\alpha
i}|^2, \label{pureosc}
\end{equation}
for astrophysical neutrinos traversing a vast distance where
$U_{\alpha i}$ and $U_{\beta i}$ are elements of the neutrino mixing
matrix. It is easily seen that $P^{\rm osc}_{\alpha\beta}=P^{\rm
osc}_{\beta\alpha}$. Because of the probability conservation, the
flux of neutrinos on the Earth also satisfies the normalization
condition given by Eq.~(\ref{norma}). We first compute $Q^{\rm osc}$
in the tribimaximal limit \cite{tbm} of neutrino mixing angles ,
i.e., $\sin^2\theta_{23}=1/2$, $\sin^2\theta_{12}=1/3$ and
$\sin^2\theta_{13}=0$. In this limit, the $\nu_{\mu}-\nu_{\tau}$
symmetry is exact. In fact, an exact $\nu_{\mu}-\nu_{\tau}$ symmetry
amounts to the condition $|U_{\mu i}|=|U_{\tau i}|$ for $i=1, 2, 3$
\cite{Balantekin:1999dx,Harrison:2002et}. This condition can be
realized by having both $\sin^2\theta_{23}=1/2$ and
$\sin\theta_{13}\cos\delta=0$, which are respected by the above
tribimaximal limit of neutrino mixing angles. Denoting $P^{\rm osc}$
in this limit as $P_0^{\rm osc}$, we have
\begin{equation}
P_0^{\rm osc}= \left(
  \begin{array}{ccc}
    5/9 & 2/9 & 2/9 \\
    2/9 & 7/18 & 7/18 \\
    2/9 & 7/18 & 7/18 \\
  \end{array}
\right). \label{p0}
\end{equation}
Since $\nu_{\mu}-\nu_{\tau}$ symmetry is exact in this case, one can
see that the second and the third rows of $P_0^{\rm osc}$ are
identical, so are the second and third columns of $P_0^{\rm osc}$.
The corresponding $Q$ matrix in this limit is given by
\begin{equation}
Q_0^{\rm osc}\equiv {\mathbf A}^{-1}P_0^{\rm osc}{\mathbf A}= \left(
  \begin{array}{ccc}
    1 & 0 & 0 \\
    0 & 0 & 0 \\
    0 & 0 & 1/3 \\
  \end{array}
\right). \label{diagonal}
\end{equation}
As expected, $Q_{0,11}^{\rm osc}=1$ and $Q_{0,12}^{\rm
osc}=Q_{0,13}^{\rm osc}=0$. Furthermore, any element in either the
second row or the second column of $Q_0^{\rm osc}$ vanishes.

We can compute the correction to $Q_0^{\rm osc}$ as neutrino mixing
parameters deviate from the tribimaximal limit. We consider such
deviations for $\theta_{13}$ and $\theta_{23}$ while keeping
$\sin^2\theta_{12}=1/3$. In this case $P^{\rm osc}=P_0^{\rm osc}+
P_1^{\rm osc}+\cdots$ where $P_1^{\rm osc}$ is the leading order
correction in powers of $\cos2\theta_{23}$ and $\sin\theta_{13}$.
One has \cite{Posc1}
\begin{equation}
P_1^{\rm osc}= \left(
  \begin{array}{ccc}
    0 & \epsilon & -\epsilon \\
    \epsilon & -\epsilon & 0 \\
    -\epsilon & 0 & \epsilon \\
  \end{array}
\right), \label{p1osc}
\end{equation}
where $\epsilon=2
\cos2\theta_{23}/9+\sqrt{2}\sin\theta_{13}\cos\delta/9$ with
$\delta$ the $CP$ phase. Taking into account
$P_1^{\rm osc}$, we obtain $Q^{\rm osc}=Q_0^{\rm osc}+ Q_1^{\rm
osc}$ with
\begin{equation}
Q_1^{\rm osc}={\mathbf A}^{-1}P_1^{\rm osc}{\mathbf A}= \left(
  \begin{array}{ccc}
    0 & 0 & 0 \\
    0 & 0 & -3\epsilon \\
    0 & -\epsilon & 0 \\
  \end{array}
\right). \label{Q1osc}
\end{equation}
Therefore $Q^{\rm osc}$ is given by
\begin{equation}
Q^{\rm osc}={\mathbf A}^{-1}P^{\rm osc}{\mathbf A}= \left(
  \begin{array}{ccc}
    1 & 0 & 0 \\
    0 & 0 & -3\epsilon \\
    0 & -\epsilon & 1/3 \\
  \end{array}
\right).
\label{Qosc}
\end{equation}
Because of the correction term $Q_1^{\rm osc}$, one can see from
Eq.~(\ref{Qosc}) that the $\nu_{\mu}-\nu_{\tau}$ symmetry is broken
since $Q_{23}^{\rm osc}$ and $Q_{32}^{\rm osc}$ are nonvanishing.
Focusing on the third row of $Q^{\rm osc}$, we obtain
$\lambda=b/3-a\epsilon$ from Eqs.~(\ref{new_basis}) and
(\ref{Qosc}).

We next consider models of neutrino decays. Flavor transitions of
astrophysical neutrinos due to effects of neutrino decays were
discussed in Ref.~\cite{Beacom:2002vi}. The simplest case of
neutrino decays is that both the heaviest and the middle mass
eigenstates decay to the lightest mass eigenstate. We first assume
the branching ratios for the above two decays are both $100\%$.
Under this condition, the transition matrix is given by $P^{\rm
dec}_{\alpha\beta}=|U_{\alpha 1}|^2$ for the normal mass hierarchy
and $P^{\rm dec}_{\alpha\beta}=|U_{\alpha 3}|^2$ for the inverted
mass hierarchy. The corresponding matrix $Q^{\rm dec}$ then reads
\begin{equation}
Q^{\rm dec}= \left(
  \begin{array}{ccc}
    1 & 0 & 0 \\
    3(|U_{\tau j}|^2-|U_{\mu j}|^2)/2 & 0 & 0 \\
    |U_{e j}|^2-(|U_{\mu j}|^2+|U_{\tau j}|^2)/2 & 0 & 0 \\
  \end{array}
\right), \label{decay_hm}
\end{equation}
where $j=1$ for the normal mass hierarchy and $j=3$ for the inverted
mass hierarchy. One can see that $Q^{\rm dec}_{11}=1$ and $Q^{\rm
dec}_{12}=Q^{\rm dec}_{13}=0$. Furthermore, in the limit of exact
$\nu_{\mu}-\nu_{\tau}$ symmetry, one has $|U_{\tau j}|=|U_{\mu j}|$
such that the elements in both the second row and the second column
of $Q^{\rm dec}$ vanish. If branching ratios for the above decays
are not $100\%$, the resulting $Q^{\rm dec}$ matrix would be
different but nevertheless gives rise to the same neutrino flavor
ratio on the Earth. It is interesting to see that all the
nonvanishing elements of $Q^{\rm dec}$ are located in the first
column. Hence, following Eq.~(\ref{new_basis}), the neutrino flavor
ratio on the Earth is independent of the neutrino flavor ratio at
the source in this scenario.

\begin{table}
  \centering
  \begin{tabular}{c|c|c|c|c}
\hline\hline
                          &   \multicolumn{4}{c}{ Elements of subleading matrices  $Q^{\prime\rm dec}_1$ and  $Q^{\prime\prime\rm dec}_1$ }  \\ \hline
     &       12     &       21            &     23            &       32          \\    \hline
$Q^{\prime\rm dec}_1$ & $-2(1-r-s)(\epsilon_1+\epsilon_2)/3$ & $-(1+r)\epsilon_1-(1+s)\epsilon_2$ & $r\epsilon_1-(1-s)\epsilon_2$ & $[s(\epsilon_1+\epsilon_2)-\epsilon_2]/3$    \\    \hline
$Q^{\prime\prime \rm dec}_1$ & $2(1-r-s)\epsilon_1/3$ & $(1+s)\epsilon_1-(r-s)\epsilon_2$ & $-\epsilon_1-2\epsilon_2$ & $-[(1+r-s)\epsilon_1+2\epsilon_2]/3$        \\    \hline
\end{tabular}
\caption{Nonzero elements for subleading matrices  $Q^{\prime\rm
dec}_1$ and $Q^{\prime\prime\rm dec}_1$. The indices $12, 21, 23$,
and $32$ in the heading of the table denote the positions of matrix
elements. $r$ and $s$ denote branching ratios for the decays $\nu_3
\to \nu_2$ and $\nu_3 \to \nu_1$, respectively, in the case of
normal mass hierarchy, and branching ratios for the decays $\nu_2
\to \nu_1$ and $\nu_2 \to \nu_3$, respectively, in the case of
inverted mass hierarchy.
$\epsilon_1\equiv(\cos2\theta_{23}-\sqrt{2}\sin\theta_{13}\cos\delta)/3$
and $\epsilon_2\equiv\cos2\theta_{23}/2-\epsilon_1$. To obtain these
expressions, we have taken $\sin^2\theta_{12}=1/3$.  }
\end{table}

Let us consider another neutrino decay scenario where only the
heaviest neutrino decays. Following earlier treatments, we set
$\sin^2\theta_{12}=1/3$ while allowing $\theta_{23}$ and
$\theta_{13}$ to deviate from $\pi/4$ and $0$, respectively. For the
normal mass hierarchy, we write $Q^{\prime \rm dec}=Q^{\prime \rm
dec}_0+Q^{\prime \rm dec}_1$ where $Q^{\prime \rm dec}_0$ is the
leading term obtained in the limit $\sin^2\theta_{23}=1/2$ and
$\sin\theta_{13}=0$, while $Q^{\prime \rm dec}_1$ is the first-order
correction which is linear in $\cos 2\theta_{23}$ and
$\sin\theta_{13}$. We find
\begin{equation}
Q^{\prime \rm dec}_0= \frac{1}{6}\left(
  \begin{array}{ccc}
    4+2(r+s) & 0 & 2-2(r+s) \\
    0 & 0 & 0 \\
    1+s & 0 & 1-s \\
  \end{array}
\right), \label{decay_h0}
\end{equation}
and
\begin{equation}
Q^{\prime \rm dec}_1= \left(
  \begin{array}{ccc}
    0 & (Q^{\prime \rm dec}_1)_{12} & 0 \\
    (Q^{\prime \rm dec}_1)_{21} & 0 & (Q^{\prime \rm dec}_1)_{23} \\
    0 &(Q^{\prime \rm dec}_1)_{32} & 0 \\
  \end{array}
\right), \label{decay_h1}
\end{equation}
where $r$ and $s$ are the branching ratios for the decay modes
$\nu_3 \to \nu_2$ and $\nu_3 \to \nu_1$ respectively. The nonzero
elements of $Q^{\prime \rm dec}_1$ are given in Table I.

If $\nu_3$ exclusively decays into either $\nu_2$ or $\nu_1$, one
has $r+s=1$. In this limit, $Q^{\prime \rm dec}_{11}=1$, $Q^{\prime
\rm dec}_{12}=Q^{\prime \rm dec}_{13}=0$ as expected. One also
observes that the elements in the second row and the second column
of the leading matrix $Q^{\prime \rm dec}_0$ vanish due to
$\nu_{\mu}-\nu_{\tau}$ symmetry. Finally, the third row of
$Q^{\prime \rm dec}$ gives rise to
$\lambda=\left[(1+3b)+(1-3b)s\right]/18+a\left[s(\epsilon_1+\epsilon_2)-\epsilon_2\right]/3$.
Focusing on the leading order contributions, one has $\lambda \geq
0$ since $b\leq 1/3$; i.e., $\phi_e$ is either equal or larger than
$(\phi_{\mu}+\phi_{\tau})/2$ irrespective of the flavor ratio at the
source. For comparison, the standard oscillation scenario gives
$\lambda= b/3$ at the leading order, which is either positive or
negative depending on the sign of $b$.

For the inverted mass hierarchy, we denote $r$ and $s$ as branching
ratios for the decay modes $\nu_2\to \nu_1$ and $\nu_2\to \nu_3$,
respectively. We obtain $Q^{\prime \prime \rm dec}=Q^{\prime \prime
\rm dec}_0+Q^{\prime \prime \rm dec}_1$ with
\begin{equation}
Q^{\prime \prime \rm dec}_0= \frac{1}{6}\left(
  \begin{array}{ccc}
    4+2(r+s) & 0 & 0 \\
    0 & 0 & 0 \\
    r-s & 0 & 2 \\
  \end{array}
\right), \label{decay_h3}
\end{equation}
and
\begin{equation}
Q^{\prime \prime \rm dec}_1= \left(
  \begin{array}{ccc}
    0 & (Q^{\prime \prime\rm dec}_1)_{12} & 0 \\
    (Q^{\prime \prime\rm dec}_1)_{21} & 0 & (Q^{\prime \prime \rm dec}_1)_{23} \\
    0 &(Q^{\prime \prime \rm dec}_1)_{32} & 0 \\
  \end{array}
\right). \label{decay_h4}
\end{equation}
The nonzero matrix elements of $Q^{\prime \prime \rm dec}_1$ are
given in Table I. In the limit $r+s=1$, $Q^{\prime\prime \rm
dec}_{11}=1$, $Q^{\prime\prime \rm dec}_{12}=Q^{\prime\prime \rm
dec}_{13}=0$ as expected. It is also observed that the elements in
the second row and the second column of the leading matrix
$Q^{\prime\prime \rm dec}_0$ vanish. Finally, the third row of
$Q^{\prime\prime \rm dec}$ gives rise to
$\lambda=(r-s+6b)/18-a[(1+r-s)\epsilon_1+2\epsilon_2]/3$.

As the last example, we discuss neutrino flavor transitions affected
by the decoherence effect from the Planck-scale physics
\cite{Lisi:2000zt}. In a three-flavor framework, it has been shown
that \cite{Gago:2002na,Hooper:2004xr,Anchordoqui:2005gj}
\begin{eqnarray}
P_{\alpha\beta}^{\rm dc}&=&\frac{1}{3}+\left[\frac{1}{2}e^{-\gamma_3 d}(U_{\beta 1}^2-U_{\beta 2}^2)(U_{\alpha 1}^2-U_{\alpha 2}^2)\right. \nonumber \\
 &+&\left.\frac{1}{6}e^{-\gamma_8 d}(U_{\beta 1}^2+U_{\beta 2}^2-2U_{\beta 3}^2)(U_{\alpha 1}^2+U_{\alpha 2}^2\right.\nonumber \\
 &-& \left. 2U_{\alpha 3}^2)\right],
\end{eqnarray}
where $\gamma_3$ and $\gamma_8$ are eigenvalues of the decoherence
matrix, and $d$ is the neutrino propagating distance from the
source. The $CP$ phase in the neutrino mixing matrix $U$ has been
set to zero. Taking $\gamma_3=\gamma_8=\gamma$, we obtain $Q^{\rm
dc}\equiv {\mathbf A}^{-1}P^{\rm dc}{\mathbf A}=Q_0^{\rm
dc}+Q_1^{\rm dc}$ where
\begin{equation}
Q^{\rm dc}_0= \left(
  \begin{array}{ccc}
    1 & 0 & 0 \\
    0 & 0 & 0 \\
    0 & 0 & e^{-\gamma d}/3 \\
  \end{array}
\right), \label{decoherence_0}
\end{equation}
and
\begin{equation}
Q^{\rm dc}_1= e^{-\gamma d}\left(
  \begin{array}{ccc}
    0 & 0 & 0 \\
    0 & 0 & -3\epsilon_0 \\
    0 & -\epsilon_0 & 0 \\
  \end{array}
\right), \label{decoherence_1}
\end{equation}
with $\epsilon_0=2 \cos2\theta_{23}/9+\sqrt{2}\sin\theta_{13}/9$.
From the definition right below Eq.~(\ref{p1osc}), we note that
$\epsilon_0=\epsilon(\delta=0)$. One can see that $Q^{\rm
dc}_{11}=1$, and $Q^{\rm dc}_{12}=Q^{\rm dc}_{13}=0$. Furthermore
the elements in the second row and the second column of the leading
matrix $Q^{\rm dc}_0$ vanish. In the absence of the decoherence
effect, i.e., $\gamma \to 0$, it is seen that $Q^{\rm dc}$ reduces
to $Q^{\rm osc}$. In the full decoherence case, i.e., $e^{-\gamma
d}\to 0$, we have $\kappa=1/3$, $\rho=\lambda=0$ such that
$\phi_e:\phi_{\mu}:\phi_{\tau}=1:1:1$.

{\it Probing $Q$ by measuring flavor ratios of astrophysical
neutrinos.}-- We have shown that the flavor transitions of
astrophysical neutrinos can be parametrized by the matrix $Q$. As we
have argued earlier, the $Q$ matrix is very convenient for
classifying flavor transition models. One could determine the matrix
elements $Q_{ij}$ by measuring flavor ratios of astrophysical
neutrinos arriving on the Earth. In this regard, we derive from
Eqs.~(\ref{new_basis}) and (\ref{measure_rl}) that
\begin{eqnarray}
 3\left(f_{\tau}(a,b)-f_{\mu} (a,b)\right)/2&=& \left(\frac{1}{3}Q_{21}+aQ_{22}+bQ_{23}\right)/\kappa (a,b),\label{measureQ2} \\
 f_e (a,b)-\left(f_{\mu} (a,b)+f_{\tau}(a,b)\right)/2&=&\left(\frac{1}{3}Q_{31}+aQ_{32}+bQ_{33}\right)/\kappa (a,b),\label{measureQ}
\end{eqnarray}
where $f_{\alpha}$ is the fraction of $\nu_{\alpha}$, i.e.,
$f_{\alpha}\equiv
\phi_{\alpha}/(\phi_e+\phi_{\mu}+\phi_{\tau})=\phi_{\alpha}/3\kappa$.
In the above equations, we have explicitly denoted the dependence of
$f_{\alpha}$ on the source parameters $a$ and $b$. Furthermore we
also indicated that $\kappa$ is generally a function of source
parameters since the total neutrino flux is not necessarily
conserved during neutrino propagations.

In the flux-conservation case, $Q_{11}=1$ and $Q_{12}=Q_{13}=0$,
which gives $\kappa=1/3$. In principle, the matrix elements $Q_{2i}$
in the second row of $Q$ can be solved from Eq.~(\ref{measureQ2}) by
inputting three sets of $f_{\alpha} (a,b)$ measured from three
different astrophysical sources. Here we assume precise knowledge of
parameters $a$ and $b$ from each source. The matrix elements
$Q_{3i}$ in the third row of $Q$ can be solved from
Eq.~(\ref{measureQ}) in a similar way. In the case that
$\nu_{\mu}-\nu_{\tau}$ symmetry is not significantly broken, one
expects $Q_{21}, Q_{22}$, and $Q_{23}$ are all suppressed.
Therefore, it is more involved to probe the second row of $Q$ than
to probe the third one. To probe the latter, we have
\begin{equation}
 f_e (a,b)/3-\left(f_{\mu} (a,b)+f_{\tau}(a,b)\right)/6\approx
 \frac{1}{3}Q_{31}+bQ_{33},
\label{3rd_probe}
\end{equation}
since $Q_{32}$ is also suppressed due to the approximate
$\nu_{\mu}-\nu_{\tau}$ symmetry. We note that $f_{\alpha} (a,b)$ on
the left-hand side of Eq.~(\ref{3rd_probe}) only depends on $b$. It
is possible to solve for $Q_{31}$ and $Q_{33}$ if the measurement on
$f_e-(f_{\mu}+f_{\tau})/2$ can be performed with respect to two
different astrophysical sources where the value of the $b$ parameter
in each source is known.

In the case of flux nonconservation, the function $\kappa (a,b)$ is
not known since it is difficult to determine the absolute flux of
astrophysical neutrinos at the source. Hence one cannot directly
solve for $Q_{2i}$ and $Q_{3i}$ from Eqs.~(\ref{measureQ2}) and
(\ref{measureQ}) by inputting $f_{\alpha} (a,b)$ from measurements.
On the other hand, the signature for $\kappa \neq 1/3$ could still
be detected by the following consistency analysis. We recall from
Eq.~(\ref{measureQ}) that the third row of $Q$ is related to the
measurement by
\begin{equation}
f_e (a,b)-\left(f_{\mu} (a,b)+f_{\tau}(a,b)\right)/2\approx
\left(\frac{1}{3}Q_{31}+bQ_{33}\right)/\kappa
(a,b).\label{kappa_neq}
\end{equation}
As it was just argued, one could set $\kappa=1/3$ in the above
equation and invoke two astrophysical sources to solve for $Q_{31}$
and $Q_{33}$. However, taking this set of $Q_{31}$ and $Q_{33}$ as
an input, one expects that the right-hand side of
Eq.~(\ref{kappa_neq}) is likely to be inconsistent with the
left-hand side obtained from the third astrophysical source.

It is clear that the knowledge of the neutrino flavor ratio at the
source is crucial for probing the matrix $Q$. Previous studies
\cite{Kashti:2005qa,Lipari:2007su} pointed out that this ratio is
energy dependent for a general astrophysical source. For parent
pions with an $E^{-2}$ energy spectrum, the flavor ratio of
neutrinos arising from the decays of these pions and the subsequent
muon decays is $\phi_{0,e}:\phi_{0,\mu}:\phi_{0,\tau}=1:1.86:0$ at
low energies \cite{kinematics} where energy losses of pions and
muons in the source are negligible. The ratio
$\phi_{0,e}/\phi_{0,\mu}$ however decreases with the increase of
muon (pion) energy and eventually approaches zero. This behavior
results from the above-mentioned energy losses which are important
at higher energies. Recently, a systematic study on possible
neutrino flavor ratios from cosmic accelerators listed on the Hillas
plot was initiated \cite{Hummer:2010ai}. The neutrino flavor ratio
at the source depends on the spectrum index of injecting protons,
the size of the acceleration region, and the magnetic field strength
at the source. In some regions of the above-mentioned parameters,
the neutrino flavor ratios are energy dependent, while in some other
parameter regions they could behave as those of a pion source or
those of a muon-damped source, which are both energy independent. In
the following, we illustrate the determination of $Q_{31}$ and
$Q_{33}$ by measuring flavor ratios of astrophysical neutrinos
arriving on the Earth from a pion source and a muon-damped source,
respectively.
\begin{figure}[htbp]
\includegraphics[scale=0.30]{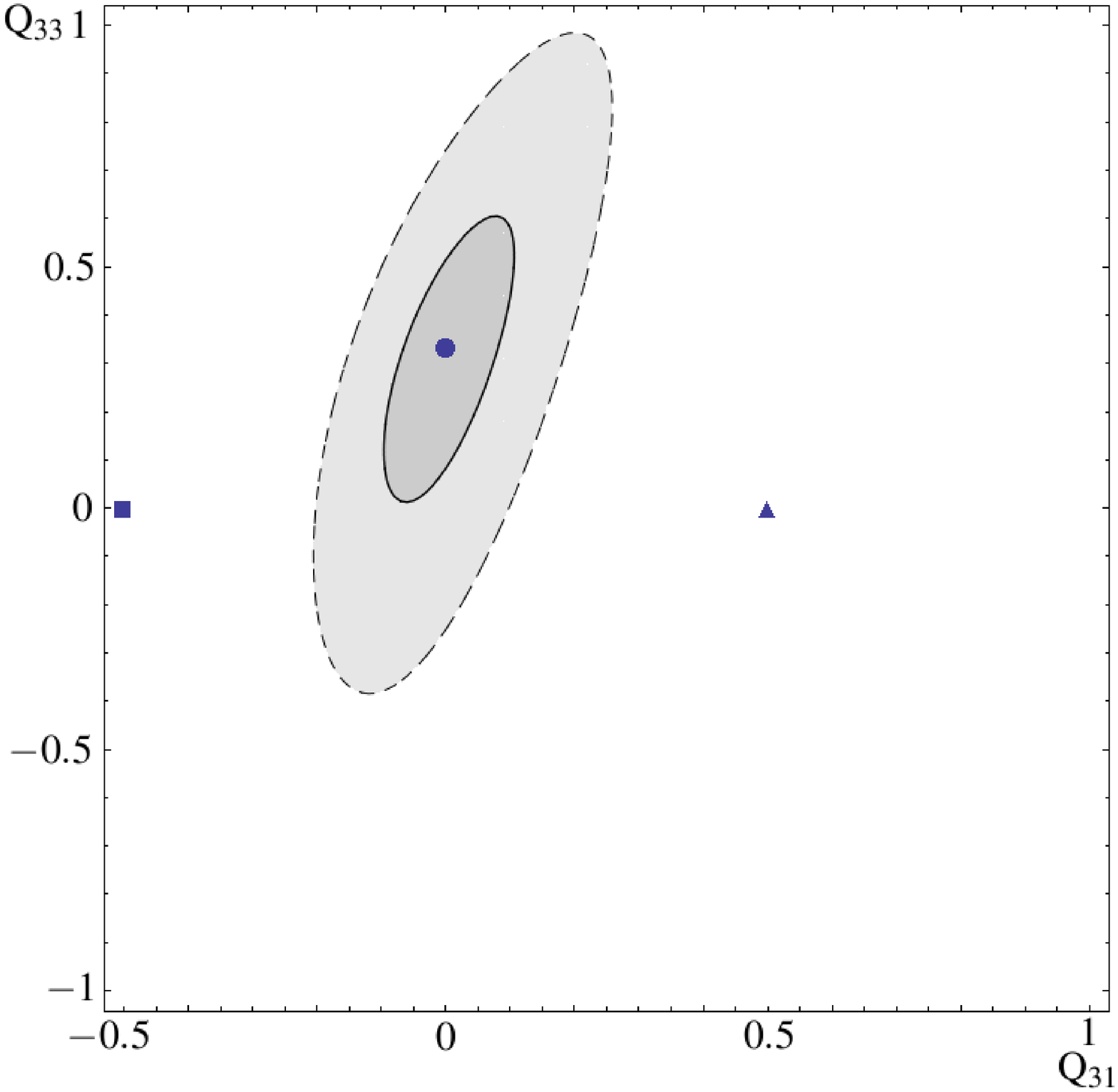}
\includegraphics[scale=0.30]{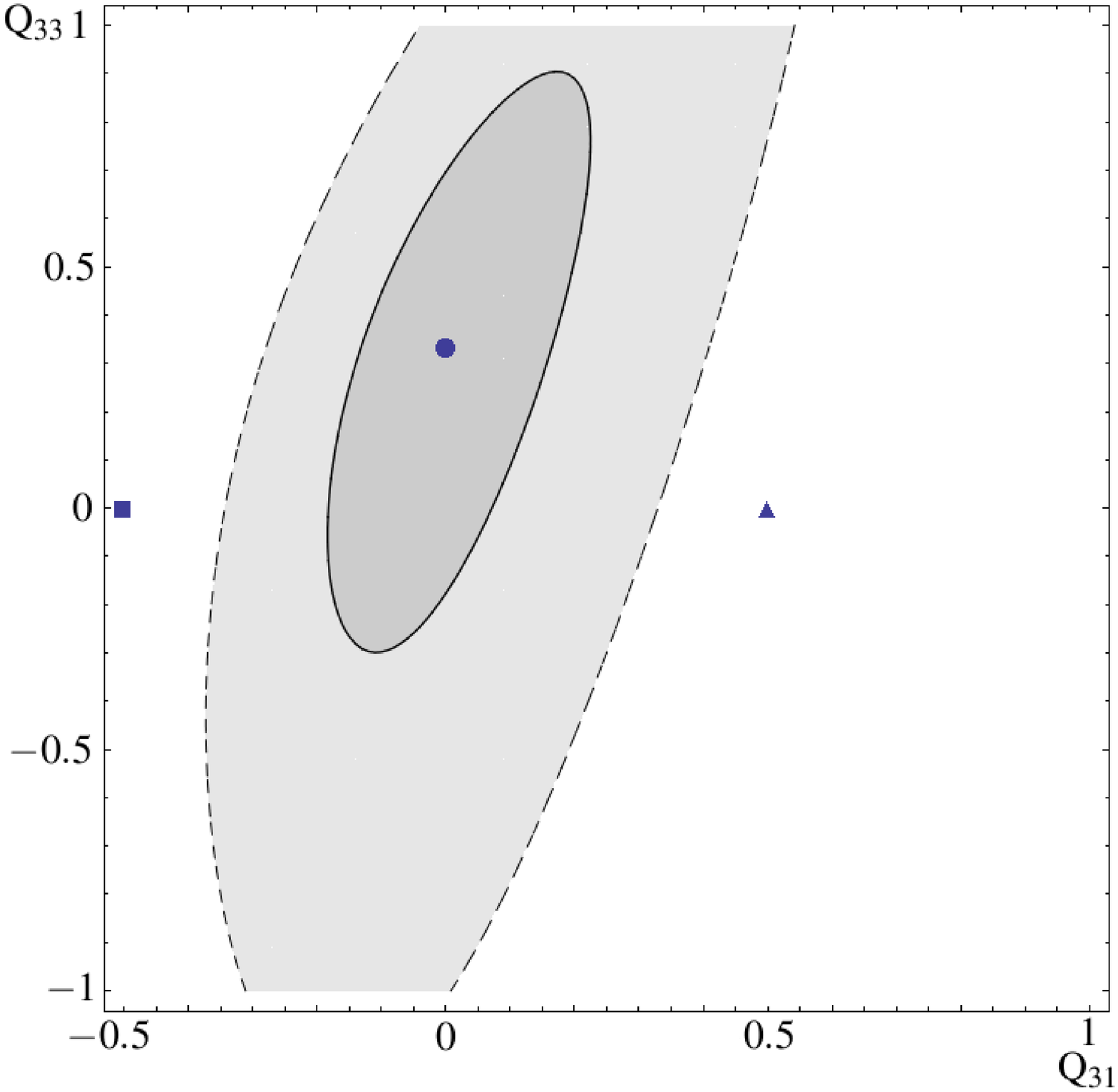}
\caption{The fitted $1\sigma$ (solid line) and $3\sigma$ (dashed
line) ranges for $Q_{31}$ and $Q_{33}$. The left panel is obtained
with measurement accuracies $\Delta R_{\pi}/R_{\pi}=\Delta
R_{\mu}/R_{\mu}=10\%$, while the right panel is obtained with
$\Delta R_{\pi}/R_{\pi}=\Delta R_{\mu}/R_{\mu}=20\%$. The circle
describes the best-fit parameter values $Q_{31}=0$ and
$Q_{33}=0.33$, corresponding to the input flavor transition model.
For reference, the parameter values for the neutrino decay scenario
given by Eq.~(\ref{decay_hm}) are denoted by the triangle and the
square, respectively, for normal and inverted mass hierarchies. The
former corresponds to $(Q_{31},Q_{33})=(0.5,0)$, while the latter
corresponds to $(Q_{31},Q_{33})=(-0.5,0)$ for neutrino mixing
parameters taking the tribimaximal values. } \label{true_osc}
\end{figure}
\begin{figure}[htbp]
\includegraphics[scale=0.30]{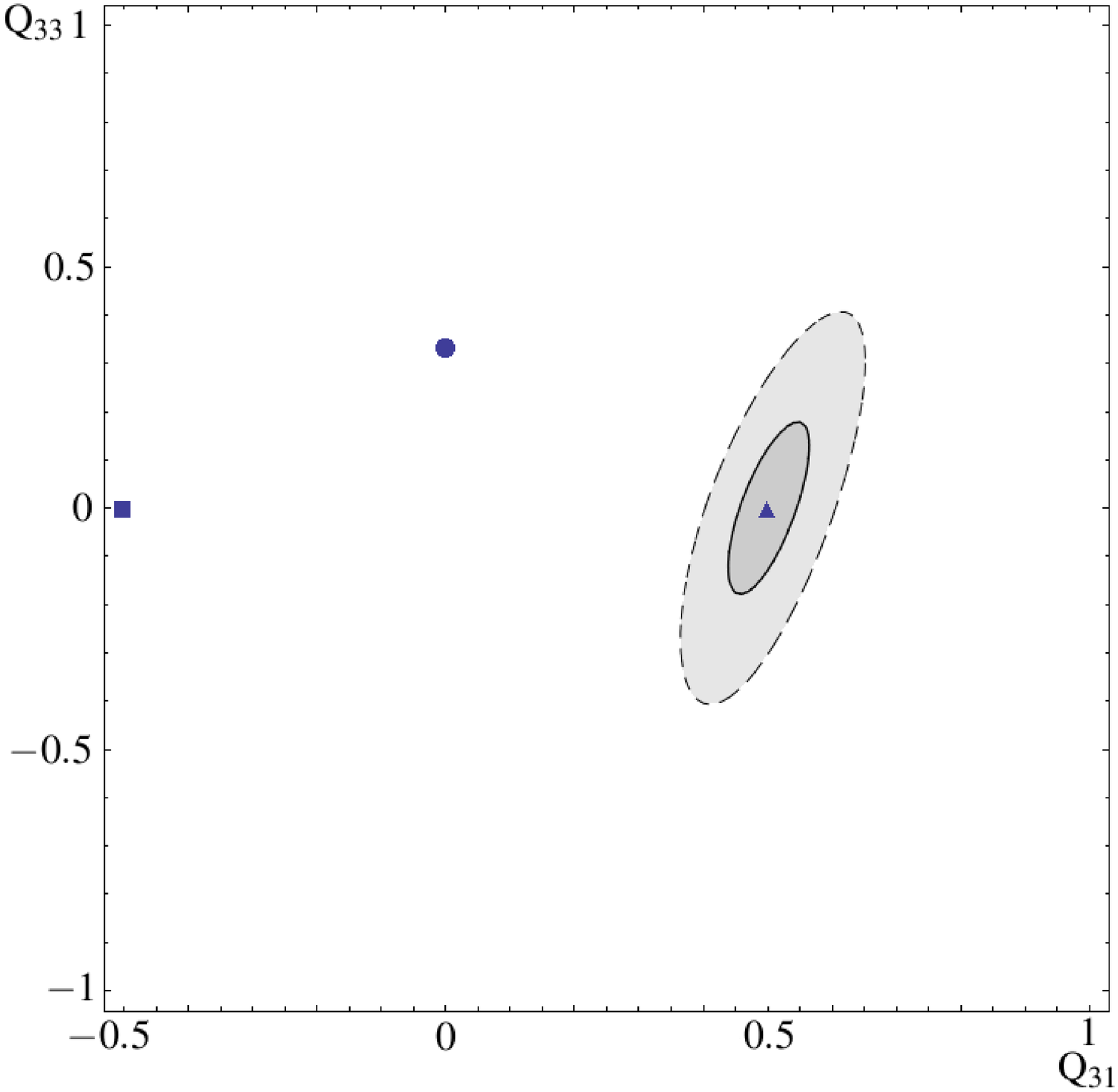}
\includegraphics[scale=0.30]{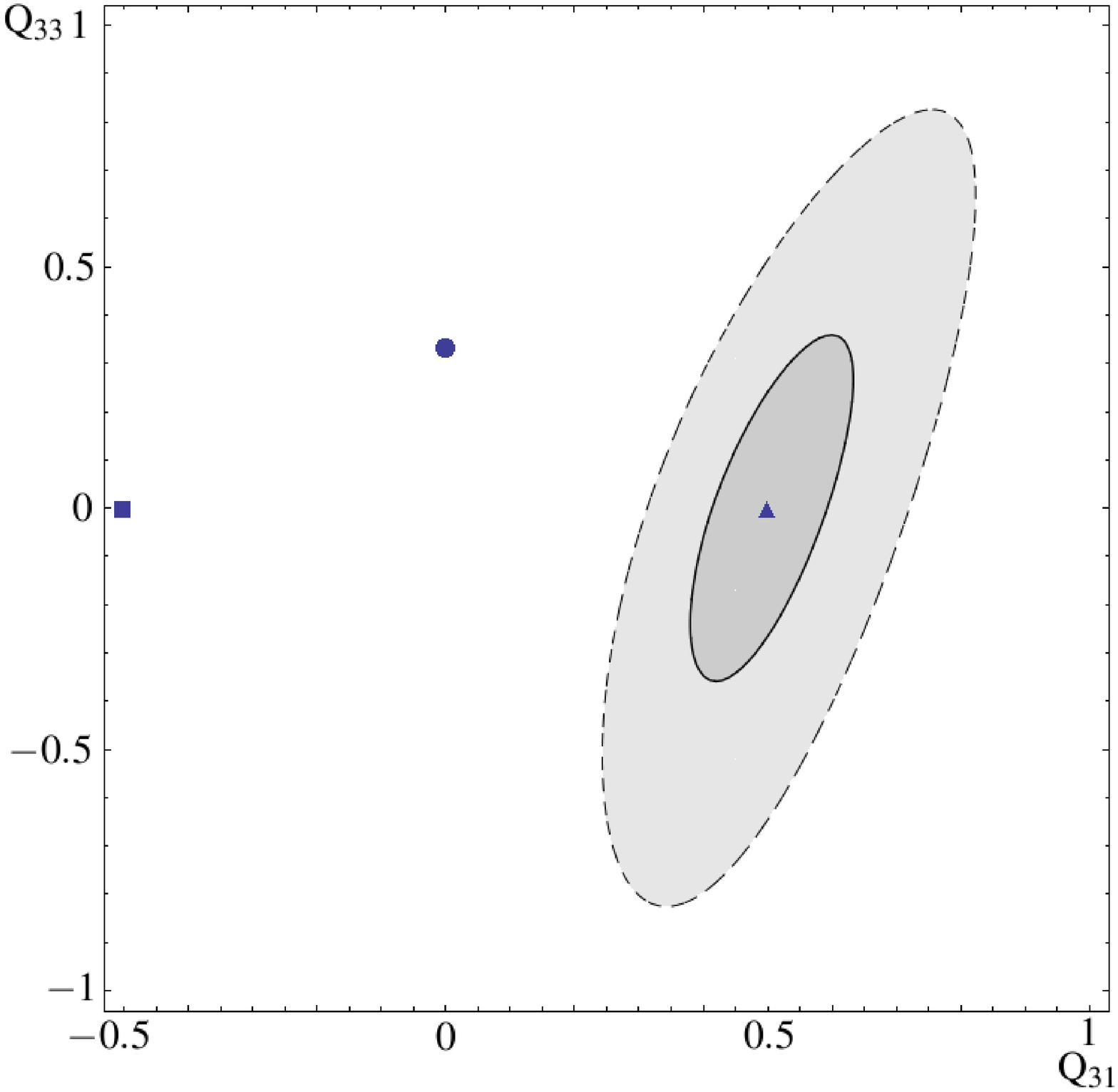}
\caption{The fitted $1\sigma$ (solid line) and $3\sigma$ (dashed
line) ranges for $Q_{31}$ and $Q_{33}$. The left panel is obtained
with measurement accuracies $\Delta R_{\pi}/R_{\pi}=\Delta
R_{\mu}/R_{\mu}=10\%$, while the right panel is obtained with
$\Delta R_{\pi}/R_{\pi}=\Delta R_{\mu}/R_{\mu}=20\%$. The triangle
describes the best-fit parameter values, $(Q_{31},Q_{33})=(0.5,0)$,
corresponding to the input flavor transition model. The circle
corresponds to the standard neutrino oscillation model, while the
square corresponds to the neutrino decay scenario given by
Eq.~(\ref{decay_hm}) with inverted mass hierarchy. }
\label{true_decay_n}
\end{figure}
\begin{figure}[htbp]
\includegraphics[scale=0.30]{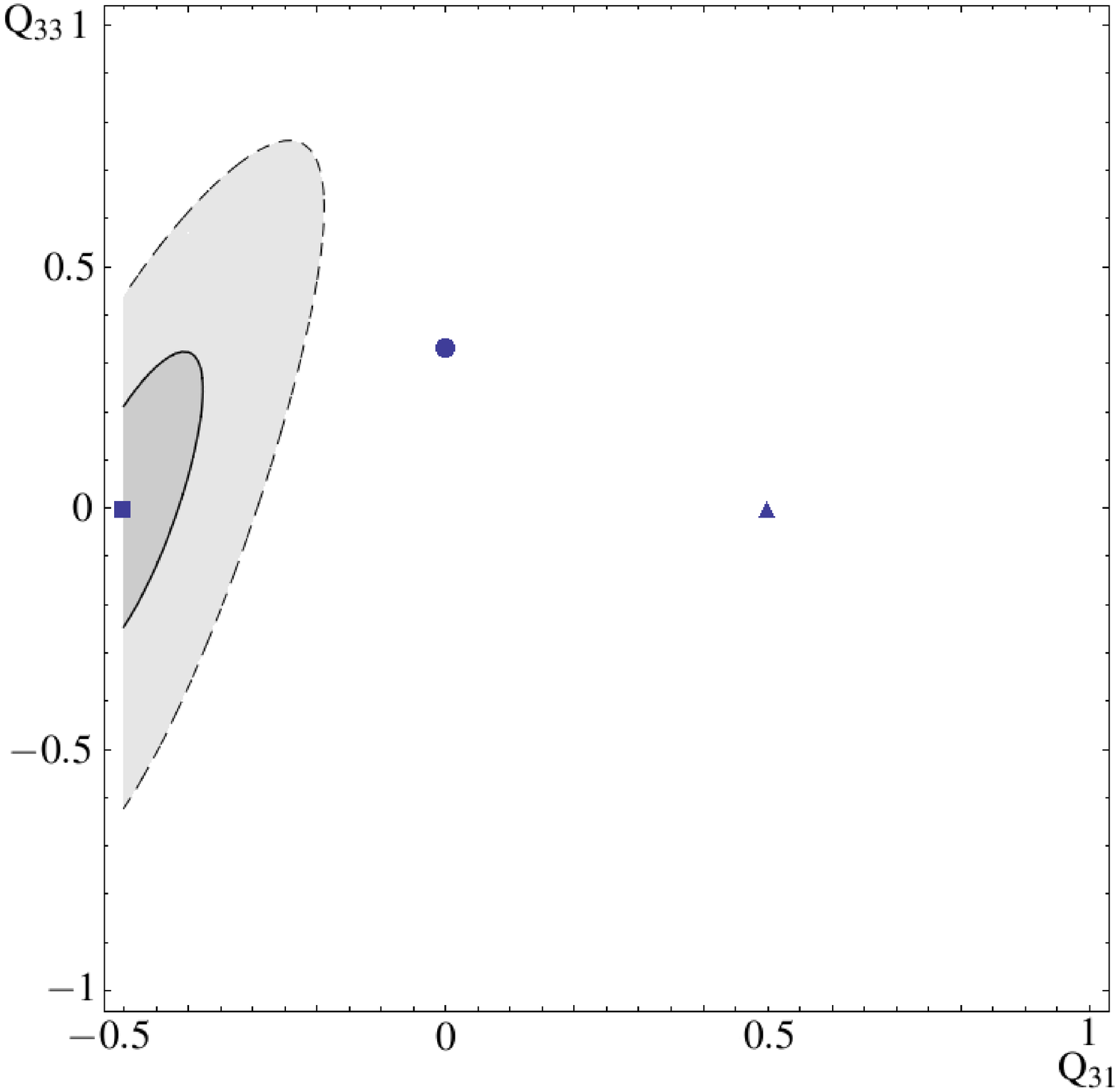}
\includegraphics[scale=0.30]{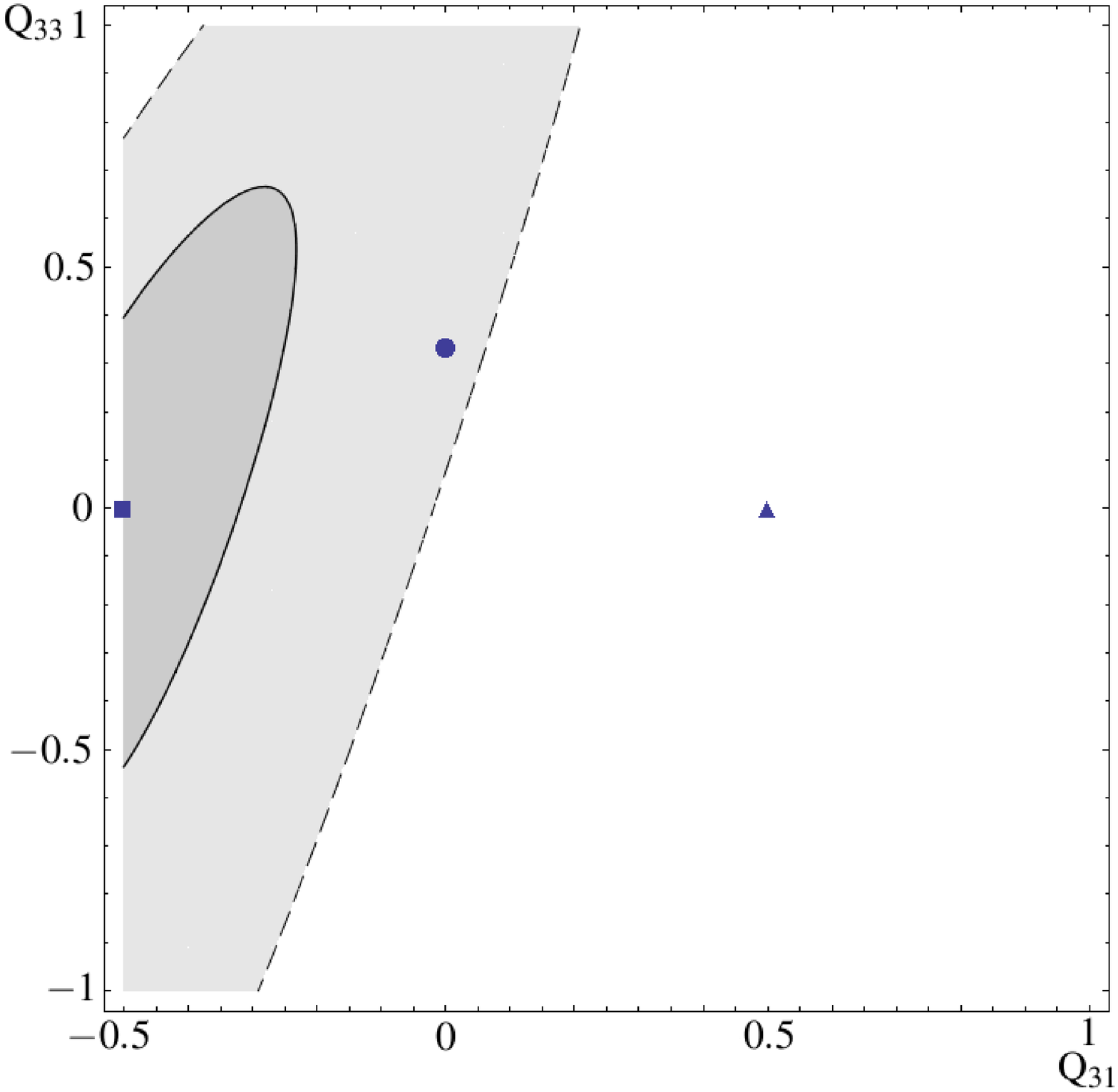}
\caption{The fitted $1\sigma$ (solid line) and $3\sigma$ (dashed
line) ranges for $Q_{31}$ and $Q_{33}$. The left panel is obtained
with measurement accuracies $\Delta R_{\pi}/R_{\pi}=\Delta
R_{\mu}/R_{\mu}=10\%$, while the right panel is obtained with
$\Delta R_{\pi}/R_{\pi}=\Delta R_{\mu}/R_{\mu}=20\%$. The square
describes the best-fit parameter values, $(Q_{31},Q_{33})=(-0.5,0)$,
corresponding to the input flavor transition model. The circle
corresponds to the standard neutrino oscillation model, while the
triangle corresponds to the neutrino decay scenario given by
Eq.~(\ref{decay_hm}) with normal mass hierarchy. }
\label{true_decay_I}
\end{figure}

To determine $Q_{31}$ and $Q_{33}$, we assume an exact
$\nu_{\mu}-\nu_{\tau}$ symmetry so that $\phi_{\mu}=\phi_{\tau}$.
The measurement of muon track to shower ratio \cite{Beacom:2003nh}
in a neutrino telescope such as IceCube can be used to extract the
flux ratio $R\equiv \phi_{\mu}/(\phi_e+\phi_{\tau})$. Clearly $R$
depends on the source parameter $b$ and the matrix elements $Q_{31}$
and $Q_{33}$ as can be seen from Eq.~(\ref{3rd_probe}). One can in
principle disentangle $Q_{31}$ and $Q_{33}$ by measuring $R$ from
two sources with different $b$ values, say a pion source with $b=0$
and a muon-damped source with $b=-1/6$. Taking into account
experimental errors in determining $R$, the ranges for $Q_{31}$ and
$Q_{33}$ in a given confidence level can be determined by the
formula
\begin{equation}
\chi^2=\left(\frac{R_{\pi,\rm th}-R_{\pi,\rm
exp}}{\sigma_{R_{\pi,\rm exp}}}\right)^2+ \left(\frac{R_{\mu,\rm
th}-R_{\mu,\rm exp}}{\sigma_{R_{\mu,\rm exp}}}\right)^2, \label{chi}
\end{equation}
where $R_{\pi,\rm exp}$ and $R_{\mu,\rm exp}$ are experimentally
measured flux ratios for neutrinos coming from a pion source and
muon-damped source, respectively, while $R_{\pi,\rm th}$ and
$R_{\mu,\rm th}$, which depend on $Q_{31}$ and $Q_{33}$, are
theoretically predicted values for $R_{\pi}$ and $R_{\mu}$
respectively. Furthermore, $\sigma_{R_{\pi,\rm exp}}=(\Delta
R_{\pi}/R_{\pi})R_{\pi,\rm exp}$ and $\sigma_{R_{\mu,\rm
exp}}=(\Delta R_{\mu}/R_{\mu})R_{\mu,\rm exp}$ with $\Delta R_{\pi}$
and $\Delta R_{\mu}$ the experimental errors in determining $R$ for
neutrinos coming from a pion source and muon-damped source,
respectively. One does not need to include uncertainties of neutrino
mixing angles $\theta_{ij}$ and $CP$ phase $\delta$ in
Eq.~(\ref{chi}) since their effects are already embedded in $Q_{31}$
and $Q_{33}$.

Let us first take the input (true) flavor transition mechanism to be
a standard neutrino oscillation model with neutrino mixing
parameters taking the tribimaximal values. One expects that
$R_{\pi,\rm exp}$ and $R_{\mu,\rm exp}$ are around $0.50$ and
$0.64$, respectively. Applying the $\chi^2$ analysis,
Eq.~(\ref{chi}), with given accuracies $\sigma_{R_{\pi,\rm exp}}$
and $\sigma_{R_{\mu,\rm exp}}$, the fitted $1\sigma$ and $3\sigma$
ranges for $Q_{31}$ and $Q_{33}$ are presented in
Fig.~\ref{true_osc}. We note that the left panel is obtained with
$\Delta R_{\pi}/R_{\pi}=\Delta R_{\mu}/R_{\mu}=10\%$ while the right
panel is the result of taking $\Delta R_{\pi}/R_{\pi}=\Delta
R_{\mu}/R_{\mu}=20\%$. For both measurement accuracies, the neutrino
decay scenario given by Eq.~(\ref{decay_hm}) can be ruled out at the
$3\sigma$ level for both mass hierarchies. We stress that the
confidence ranges in Fig.~\ref{true_osc} can be used to test any
model with specific values for $Q_{31}$ and $Q_{33}$.

We next consider the case where the input flavor transition model is
the neutrino decay scenario given by Eq.~(\ref{decay_hm}) with
normal mass hierarchy. This model corresponds to
$(Q_{31},Q_{33})=(0.5,0)$ for neutrino mixing parameters taking the
tribimaximal values. Hence one expects that $R_{\pi,\rm exp}$ and
$R_{\mu,\rm exp}$ are both around $0.2$. Applying the $\chi^2$
analysis, we obtain the fitted $1\sigma$ and $3\sigma$ ranges for
$Q_{31}$ and $Q_{33}$ as shown in Fig.~\ref{true_decay_n}. Once
more, the left panel is obtained with $\Delta R_{\pi}/R_{\pi}=\Delta
R_{\mu}/R_{\mu}=10\%$, while the right panel results from $\Delta
R_{\pi}/R_{\pi}=\Delta R_{\mu}/R_{\mu}=20\%$. For both cases, it is
seen that the standard neutrino oscillation model and the neutrino
decay model with $(Q_{31},Q_{33})=(-0.5,0)$ (inverted mass
hierarchy) can be ruled out at the $3\sigma$ level.

Finally, if the input flavor transition model is the neutrino decay
scenario given by Eq.~(\ref{decay_hm}) with inverted mass hierarchy,
i.e., $(Q_{31},Q_{33})=(-0.5,0)$, one expects that $R_{\pi,\rm exp}$
and $R_{\mu,\rm exp}$ are both around $1.0$. Applying the $\chi^2$
analysis, we obtain the fitted $1\sigma$ and $3\sigma$ ranges for
$Q_{31}$ and $Q_{33}$ as shown in Fig.~\ref{true_decay_I}. For
$\Delta R_{\pi}/R_{\pi}=\Delta R_{\mu}/R_{\mu}=10\%$ (left panel),
it is seen that the other two models displayed on the figure can be
ruled out at the $3\sigma$ level. However, for $\Delta
R_{\pi}/R_{\pi}=\Delta R_{\mu}/R_{\mu}=20\%$ (right panel), the
standard neutrino oscillation model cannot be ruled out at the same
confidence level.

{\it Conclusion.}--In summary, we have proposed to parametrize the
flavor transitions of propagating astrophysical neutrinos by the
matrix $Q$, which is related to the usual flavor transition matrix
$P$ by $Q={\mathbf A}^{-1}P{\mathbf A}$ where ${\mathbf A}$ is given
by Eq.~(\ref{evector}). We have argued that it is much easier to
classify flavor transition models by the $Q$ matrix parametrization,
where each row of $Q$ carries a clear physical meaning as
illustrated by Eq.~(\ref{new_basis}). We have also argued that the
signature for flux nonconservation might be detectable if it is
possible to observe sufficient numbers of astrophysical neutrino
sources with different flavor ratios. For the case of flux
conservation, the above observations can probe the second and the
third rows of matrix $Q$ in a model independent fashion.

For illustration, we considered the determination of the $Q$ matrix
in the exact $\nu_{\mu}-\nu_{\tau}$ symmetry limit. The relevant
matrix elements in this case are $Q_{31}$ and $Q_{33}$. We proposed
to determine them by measuring the flux ratio $R\equiv
\phi_{\mu}/(\phi_e+\phi_{\tau})$ for astrophysical neutrinos coming
from a pion source and those coming from a muon-damped source
respectively. We fitted $Q_{31}$ and $Q_{33}$ to the measured flux
ratios  $R_{\pi,\rm exp}$ and $R_{\mu,\rm exp}$ using
Eq.~(\ref{chi}). The ranges for $Q_{31}$ and $Q_{33}$ are presented
up to the $3\sigma$ confidence level for three different input
models for neutrino flavor transitions. We have found that the
measurement accuracies $\Delta R_{\pi}/R_{\pi}=\Delta
R_{\mu}/R_{\mu}=10\%$ are sufficient to discriminate among the
standard neutrino oscillation model and neutrino decay scenario
given by Eq.~(\ref{decay_hm}) for normal and inverted mass
hierarchies. We reiterate that the confidence ranges in
Figs.~\ref{true_osc}, \ref{true_decay_n} and \ref{true_decay_I} can
be used to test any flavor transition model with specific values for
$Q_{31}$ and $Q_{33}$.

Taking a neutrino source flux $E_{\nu_e}^2{\mbox d}N_{\nu_e}/{\mbox
d}E_{\nu_e}=0.5E_{\nu_{\mu}}^2{\mbox d}N_{\nu_{\mu}}/{\mbox
d}E_{\nu_{\mu}}=10^{-7}$ GeV cm$^{-2}$ s$^{-1}$, which is roughly
the order of the Waxman-Bahcall bound \cite{Waxman:1998yy}, the
accuracy $\Delta R/R=10\%$ is reachable by a decade of data taking
in Icecube \cite{Beacom:2003nh}, as stated in the beginning of this
article. However, we stress that the Waxman-Bahcall bound is for
diffuse neutrino flux. The flux from an individual point source is
smaller. Hence it could take more than a decade to reach a $10\%$
accurate measurement on $R$ arising from a point source. The radio
extension of IceCube \cite{Allison:2009rz} is expected to accumulate
neutrino events at a much faster pace. It is crucial to study the
efficiency of flavor identification in this type of detector.

In this work, the $Q$ matrix is probed by assuming an exact
$\nu_{\mu}-\nu_{\tau}$ symmetry and a precise knowledge of the
neutrino flavor ratio at the source. Away from the
$\nu_{\mu}-\nu_{\tau}$ symmetry limit, the second row of $Q$ and
$Q_{32}$ shall become relevant in addition to $Q_{31}$ and $Q_{33}$.
Furthermore, the statistical analysis outlined by Eq.~(\ref{chi})
should be refined once the uncertainty of the neutrino flavor ratio
at the source is taken into account. We shall address these issues
in a future publication.

\noindent{\bf Acknowledgements} This work is supported by the
National Science Council of Taiwan under Grants No.
97-2811-M-009-029 and 96-2112-M-009-023-MY3, and Focus Group on
Cosmology and Particle Astrophysics, National Center for Theoretical
Sciences, Taiwan.


\begin{thebibliography}{99}
\bibitem{remark}
It is more convenient to treat $\Phi_0$ as a column vector.
\bibitem{kinematics}
In Fig.~\ref{example}, we take this ratio as $1:2:0$ by
neglecting the kinematic difference between the pion decay and the subsequent muon decay.
\end{thebibliography}
\end{document}